\documentclass[a4paper,11pt]{article}
\pdfoutput=1 % if your are submitting a pdflatex (i.e. if you have
\usepackage{jcappub} % for details on the use of the package, please
\usepackage[T1]{fontenc} % if needed
\usepackage{graphicx}
\usepackage{amsmath}

\def\aj{AJ}

\title {\boldmath Tomography of stellar halos: what does anisotropy in a stellar halo tell us?}

\author[a]{Biswajit Pandey}

\affiliation[a]{Department of Physics, Visva-Bharati University, Santiniketan, 731235, India}

\emailAdd{biswap@visva-bharati.ac.in} 

\abstract{The stellar halo of the Milky Way is known to have a highly
  lumpy structure due to the presence of tidal debris and streams
  accreted from the satellite galaxies. The abundance and distribution
  of these substructures can provide a wealth of information on the
  assembly history of the Milky Way. We use some information-theoretic
  measures to study the anisotropy in a set of Milky Way-sized stellar
  halos from the Bullock \& Johnston suite of simulations that uses a
  hybrid approach coupling semi-analytic and N-body techniques. Our
  analysis shows that the whole-sky anisotropy in each stellar halo
  increases with the distance from its centre and eventually plateaus
  out beyond a certain radius. All the stellar halos have a very
  smooth structure within a radius of $\sim 50$ kpc and a highly
  anisotropic structure in the outskirts. At a given radius, the
  anisotropies at a fixed polar or azimuthal angle have two distinct
  components: (i) an approximately isotropic component and (ii) a
  component with large density fluctuations on small spatial
  scales. We remove the contributions of the substructures and any
  non-spherical shape of the halo by randomizing the polar and
  azimuthal coordinates of the stellar particles while keeping their
  radial distances fixed. We observe that the fluctuating part of the
  anisotropy is completely eliminated, and the approximately uniform
  component of the anisotropy is significantly reduced after the
  sphericalization. A comparison between the original halos and their
  sphericalized versions reveals that the approximately uniform part
  of the anisotropy originates from the discreteness noise and the
  non-spherical shape of the halo whereas the substructures contribute
  to the fluctuating part. We show that such distinction between the
  anisotropies has the potential to constrain the shape of the stellar
  halo and its substructures.}

\begin{document}
\maketitle
\flushbottom

%\begin{keywords}
%methods: statistical - data analysis - galaxies: formation - evolution
%- cosmology: large scale structure of the Universe.
%\end{keywords}

\section{Introduction}

Understanding the formation and the evolution of galaxies remains one
of the most interesting and challenging problems in cosmology.  We do
not yet completely understand the details of galaxy formation despite
the many successes of the $\Lambda$CDM model on large scales. In the
current paradigm, the galaxies form at the centre of the dark matter
halos via cooling and condensation of baryons \citep{whiterees}. The
galaxies grow further by accretion of matter and mergers with other
galaxies. In the hierarchical clustering scenario \citep{peebles70,
  press74}, the larger structures are formed via the merger of smaller
structures, leading to even larger structures like clusters and
superclusters in the Universe. However, it is essential to test the
hierarchical growth of structures on the scales of galaxies to
understand their origin and evolution.

The Milky Way is believed to have formed via a series of mergers with
other galaxies. Our galaxy has several distinct components: the thin
and thick disk, the bulge, the bar, and a diffuse stellar halo. The
stellar halo, a quasi-spherical distribution of stars, envelops the
disk and the bulge and extends to a few hundreds of kpc. It contains
$\sim 1\%$ of the total stars in the Milky Way. The structure of the
stellar halo is intimately linked to the formation of our galaxy
\citep{freeman02}. In the case of hierarchical growth, a significant
fraction of the stellar halo is expected to have formed from the
accreted and disrupted satellite galaxies \citep{johnston96,
  helmiwhite99, bullock01, bullock05, cooper10, font11}. Consequently,
the stellar halo would serve as a repository of merger debris and
contain the fossil evidence of the accretion and merger events that
led to the formation of the galaxy. A detailed review of the Milky
Way's stellar halo can be found in \citep{helmireview}.

The classic paper by Eggen et al. (1962) \citep{eggen62} analyze the
proper motions and radial velocities of $221$ dwarf stars in the
stellar halo and find that the stars with lower metallicity tend to
move in highly eccentric orbits. They interpret this correlation as a
consequence of the formation of the metal-deficient stars in a rapid
collapse of a spherical cloud in the radial direction. Contrary to
this, Searle \& Zinn (1978) \citep{searle78} find a wide range of
metallicities in a sample of 19 globular clusters located at different
galactocentric distances. Their analysis suggests a gradual
agglomeration of the stellar halo from the accretion of many dwarf
satellites. Subsequently, many observational studies are carried out
to understand the formation of the Milky Way and its stellar halo
\citep{yanny00, majewski03, bell08, juric08, bell10, depropris10,
  sesar11, deason11, deason14, slater14, kafle14, xue15, slater16,
  belokurov18, helmi18, li19, bird19, naidu20}. Nevertheless, whether
the stars in the stellar halo form in situ in the early phases of the
collapse of the Milky Way or form in satellite galaxies that were
accreted by the Milky Way at a later stage is still a matter of
debate. The two scenarios can be tested by studying the structure of
the stellar halo \citep{majewski93}. The actual process may be a
combination of these two pictures \citep{chiba00}. The stellar halo is
expected to have ellipsoidal symmetry and devoid of substructure if
the halo stars form in situ via the radial collapse of a protogalactic
cloud many dynamical times ago. On the other hand, a significant
fraction of the stellar halo would be made up of accumulated debris if
the halo stars are accreted from the satellite galaxies. The accreted
debris from the disrupted satellite galaxies would gradually disperse
in real space. The stellar halo may appear smooth despite undergoing
many episodes of accretion. Nonetheless, debris accretd in the last
few gigayears can remain spatially coherent \citep{bullock01,
  bullock05}. Further, some information about the initial conditions
can still be recovered from the phase space distribution of the stars
\citep{helmi99}.

The identifications of the substructures and their distributions
inside the stellar halo can unravel the assembly history of the Milky
Way. So, despite its low luminosity and density, the stellar halo can
provide important constraints on scenarios for the formation and
evolution of the Galaxy. Several substructures have been identified in
the stellar halo of the Milky Way in the last few decades. The Orphan
Stream \citep{grillmair06, belokurov07a, newberg10}, the Sagittarius
dwarf tidal stream \citep{ibata94, ivezi00, yanny00}, the low-latitude
stream \citep{yanny03, ibata03}, the Virgo, Hercules–Aquila and Pisces
overdensities \citep{duffau06, juric08, belokurov07b, watkins09} are
some of the prominent substructures observed in the stellar halo of
the Milky Way. Moreover, the existence of substructures is also
confirmed in the stellar halo of M31 \citep{ibata01, ferguson02,
  zucker04, ibata07, mcconnachie09}, M33 \citep{ibata07,
  mcconnachie09} and in other nearby spirals \citep{shang98,
  delgado10, mouhcine10}. These observations indicate that accretion
from satellite galaxies definitely has some role in building the
stellar halo of a galaxy.

We also require a theoretical understanding of the formation of
stellar halos in a cosmological context. Significant efforts have been
made in this direction in the last two decades \citep{bullock01,
  bullock05, abadi06, delucia08, cooper10}. Bullock \& Johnston (2005)
\citep{bullock05} study the formation of stellar halos using a
semi-analytic approach coupled with N-body that follows accretion of
satellite galaxies onto Milky Way–sized halos with an analytic,
time-dependent rigid potential. The analytic potential in such model
includes terms for both a dark matter halo and a central baryonic
disk. Cooper et al. (2010) \citep{cooper10} follow the growth of
stellar halos by combining semi-analytic galaxy formation model
(GALFORM) with cosmological dark matter simulations from the Aquarius
project \citep{springel08}. More recently, several high-resolution
hydrodynamical simulations, such as Auriga \citep{monachesi19} and
Artemis \citep{font20}, have modelled the assembly of Milky Way-like
stellar halos. An important distinction between the Bullock \&
Johnston simulation and these simulations is that they are dynamically
self-consistent.  The model stellar halos derived from these
simulations allow one to make quantitative predictions for the density
profile, degree of substructure and the properties of the stellar
content in the halo.

The distributions of stars in the stellar halo can be highly
anisotropic due to the presence of the remnant stellar streams and
debris. The word `anisotropy' in the present work refers to the
deviations from an isotropic density distribution, measured over cells
projected on the sky. Previously, counts-in-cells measures have been
used to study the substructure-induced non-uniformity in stellar halos
\citep{bell08, helmi11}. The variance of the count in cells in such
studies can be useful measure of non-uniformity. However, we define
our anisotropy measures based on the information entropy
\citep{shannon48}. The information entropy was originally introduced
by Claude Shannon in his 1948 seminal paper "A Mathematical Theory of
Communication" \citep{shannon48}. It was proposed to quantify the
information loss during communication through noisy channels.  The
Shannon entropy $H(X)$ for any discrete random variable $X$ with $n$
outcomes $\{x_{i}:i=1,....n\}$ is defined as,
\begin{equation}
H(X) =  - \sum^{n}_{i=1} \, p(x_{i}) \, \log \, p(x_{i})
\label{eq:shannon1}
\end{equation}
where $p(x_{i})$ is the probability of the $i^{th}$ outcome and $-
\log p(x_{i})$ is the information contained in the $i^{th}$ outcome of
the random variable $X$. The negative sign ensures that the
information is always positive or zero. A certain event with a
probability of 1 would contain no information. Rare events with
smaller probability are more uncertain and contain more
information. Thus, the information entropy $H(X)$ is the average
information required to describe the random variable $X$. In other
words, it characterizes the uncertainty in the random variable.

We use the information entropy to define a set of anisotropy measures
that can quantify the variation between the counts in cells across
patches of a given solid angle on the sky. A randomly chosen star can
reside in only one of these cells. So there are as many possible
outcomes of the event as the total number of cells. The probability of
finding the aforementioned star in a given cell can be determined from
the star count in it. This probability can be used to define an
anisotropy measure by combining the information entropy and its
maximum value. A set of such anisotropy measures can be constructed,
following different strategy for grouping the cells. One can study the
variation in the cells across the entire sky or restrict the analysis
to cells in a given range of polar or azimuthal angle. The anisotropy
measure can be either a function of the radial co-ordinate or a
combination of radial, polar and azimuthal co-ordinates based on the
strategy adopted for grouping the cells.

Our choice for the information entropy based measures over the
conventional measures like variance is based on the fact that the
information entropy is related to the higher-order moments of a
probability distribution \citep{pandey16b}. Hence it can contain more
information about the distribution.

The anisotropy in the stellar halo may contain important information
about its formation. The debris from the earliest merger events is
expected to be concentrated at the center of the halo, where the
dynamical time is shorter, thus to be more thoroughly mixed in phase
space. A study of the radial variation of the full sky anisotropy in
the distribution of stars can verify this paradigm. The density
profile of the stellar halo is known to roughly scale as $r^{-3}$ in
the inner region, and a steeper variation is observed in the outer
parts \citep{bell08, deason11, slater16}. The shape of the stellar
halo can also provide useful information about its formation. The
deviations of the halo shape from the spherical symmetry may have
multiple origin. The large-scale departure from spherical symmetry may
arise due to contributions from a disk-like component
\citep{zolotov09, purcell10, font11, elias18}, dominant merger events
\citep{belokurov18, iorio19, nelson19, wu22, rey22, hahn22}, violent
relaxation \citep{tremaine99, deason13} or filamentary accretion
\citep{zentner, mandelkar}. On the other hand, the small-scale
anisotropy in a stellar halo is primarily caused by coherent
substructures \citep{bell08, xue11, naidu20}.

We intend to analyze the anisotropies in the simulated stellar halos
to infer useful information about the distribution of the
substructures and the shape of the halo. We would like to clarify here
that throughout this work, both the diffuse streams or clouds and the
bound satellites are considered as substructures. Pandey (2016)
\citep{pandey16a} propose a method for quantifying the anisotropy in a
three-dimensional distribution using information
entropy. Subsequently, the method has been used to test the isotropy
in the galaxy distribution in the 2MASS and SDSS \citep{pandey17a,
  sarkar19}. Pandey (2017) \citep{pandey17b} show that the method can
also be used to determine the linear bias parameter which describes
the relationship between the galaxy distribution and the cosmic mass
density field. Here, by applying the same technique, we obtain
anisotropy profiles for the simulated stellar halos, measured over the
whole sky and in bins of azimuthal and polar angle. We use these to
investigate the connection between potentially observable anisotropy
signals and the formation history of stellar halos. We examine the
separate contributions to the anisotropy from substructures and
large-scale variations in shape. Our results provide a theoretical
foundation for future applications of this technique to data from
surveys of the real Milky Way halo.

The paper is organised as follows. We describe the data and the method
of analysis in Section 2, discuss the results in Section 3 and present
our conclusions in section 4.

\section{Data and method of analysis}
\label{sec:datamethod}

\subsection{Stellar halo catalogue}
\label{sec:halocat}
 Bullock \& Johnston (2005) \citep{bullock05} use a semi-analytic
 formalism combined with a N-body approach to follow the dynamical
 evolution of the accreted satellite galaxies and provide the
 accretion histories for 11 Milky Way-sized halos. They assumes a
 $\Lambda$CDM framework with $\Omega_m=0.3$, $\Omega_{\Lambda}=0.7$,
 $\Omega_b\,h^2=0.024$, $h=0.7$ and $\sigma_{8}=0.9$. We use the $11$
 stellar halos from the Bullock \& Johnston (2005) suite of
 simulations for our analysis.

We want to calculate the anisotropies in the distribution of stars in
these halos in the radial, polar and azimuthal directions.

%\subsection{Mock spherical halos with smooth density profile}
%\label{sec:smoothp}
%We aim to compare the anisotropies in each stellar halo against a set
%of mock spherical halos with smooth density profile.

%We generate $10$ mock halos corresponding to each stellar halo. In
%each case, we generate spherical distribution of stars with radially
%varying density ($\sim r^{-3}$) from the centre. Each mock smooth halo
%contains the same number of stars within a given radius as the
%original stellar halo. We use a random Monte Carlo dart-board
%technique \citep{pandey13} to simulate the mock halos.

\begin{figure*}[htbp!]
\centering \includegraphics[width=7.5cm]{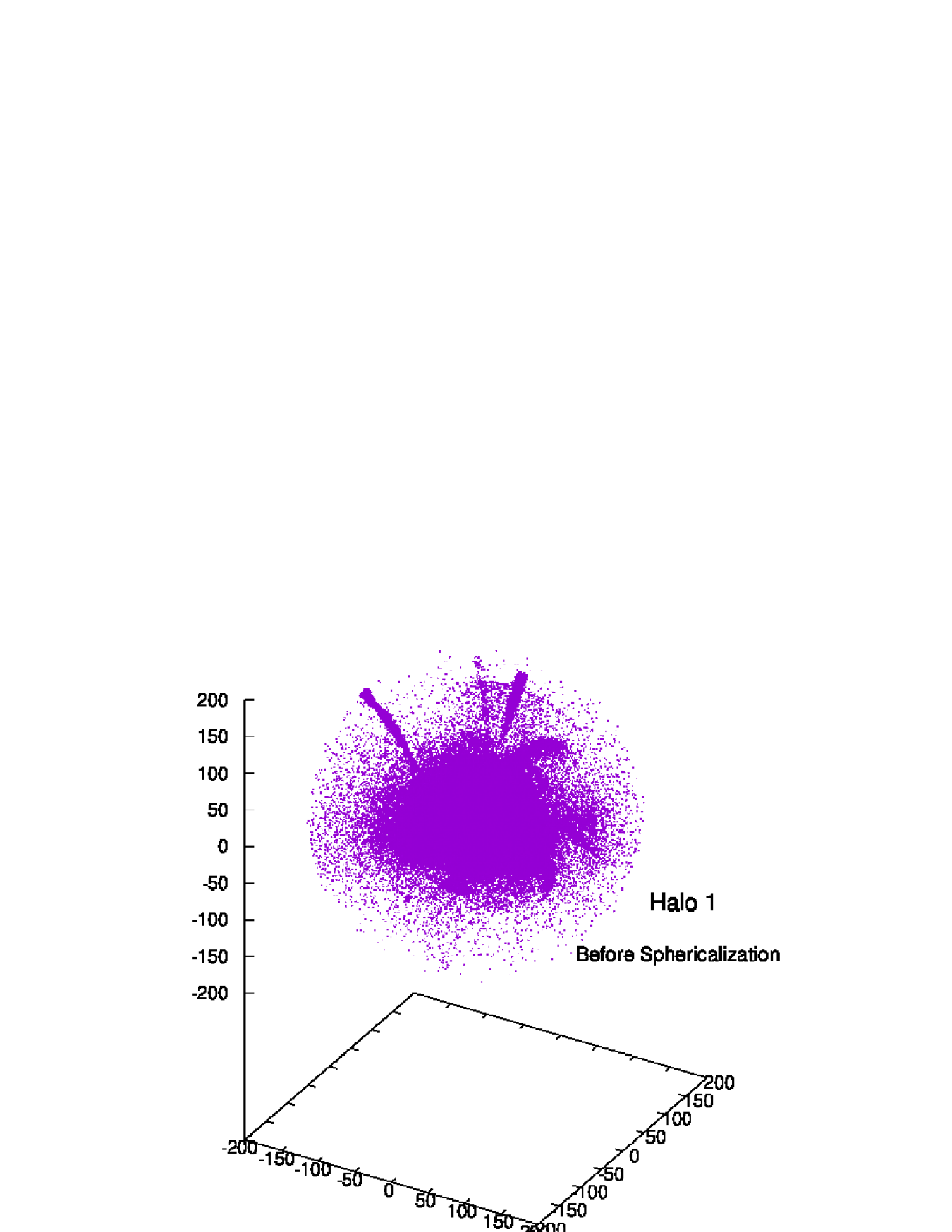}
\vspace{-0.4cm}
\includegraphics[width=7.5cm]{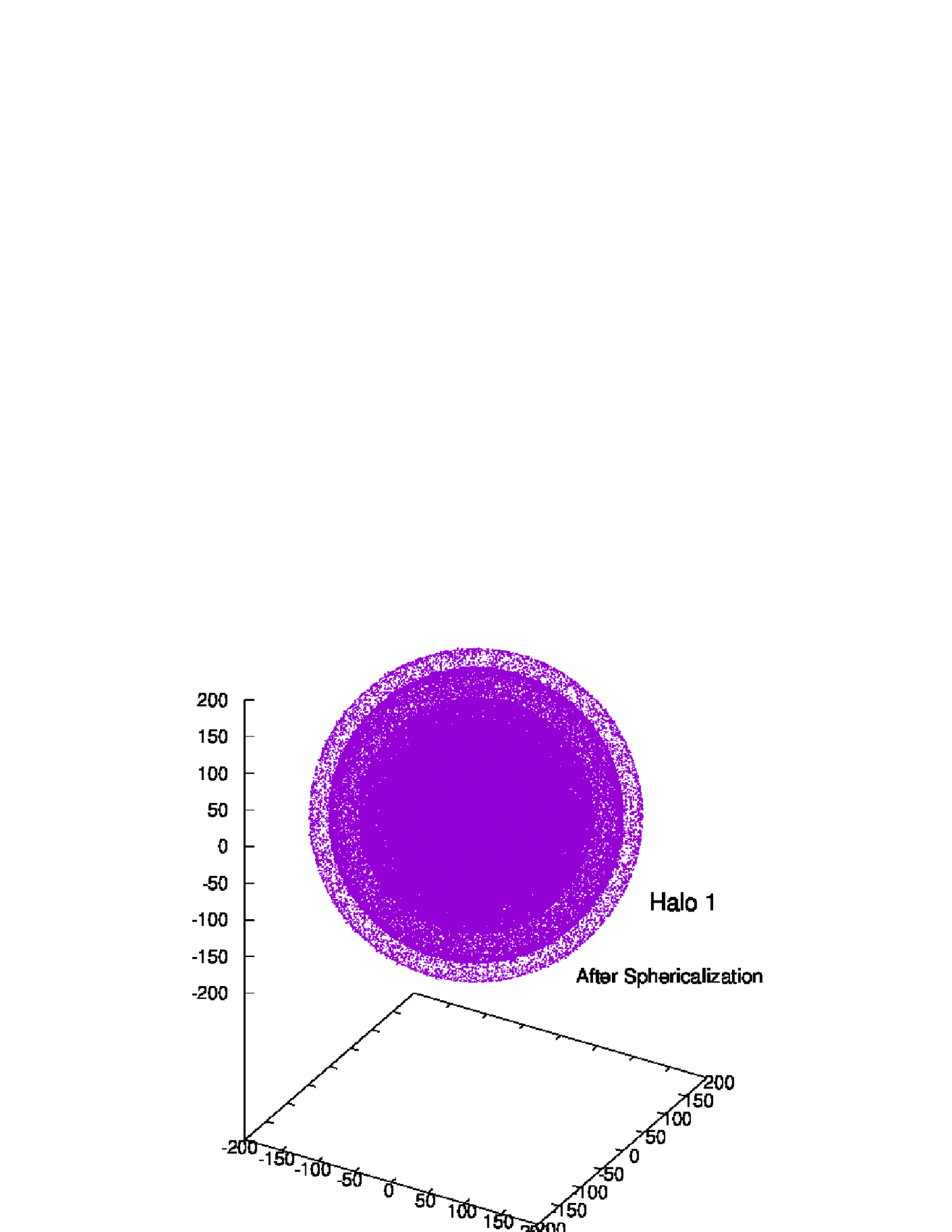}
\caption{The left panel of this figure shows the distribution of the
  stellar particles within a radius of 200 kpc. The right panel shows
  the same after the polar and the azimuthal coordinates of all the
  stellar particles are randomized while keeping their radial
  coordinates fixed.}
\label{fig:sph}
\end{figure*}

\subsection{Sphericalized halos}
\label{sec:randomize}
The stellar halos are known to be highly lumpy due to the presence of
substructures. Lumpiness in a stellar halo is expected to introduce
larger values of the anisotropy parameter. The non-spherical shape of
the stellar halo may also contribute to the measured anisotropy. We
want to quantify the anisotropies induced by the shape of the stellar
halo and the substructures within it. It would be interesting if the
contributions from the shape and substructures can be separated. It
would allow us to constrain the shape and substructures of a halo. To
investigate this possibility, we randomly extract one million stellar
particles within a radius of $200$ kpc from the centre of each stellar
halo. We prepare $10$ such randomly extracted versions for each
stellar halo. In each such halo, we then randomize the polar and
azimuthal coordinates $(\theta, \phi)$ of all the stellar particles
without changing their radial coordinates $r$
(\autoref{fig:sph}). This randomization of the $(\theta, \phi)$
coordinates would destroy all the substructures present within the
original stellar halo. Additionally, it would also obliterate all the
information about the shape of the stellar halo. It may be noted that
the radial density profile of the original stellar halo would remain
intact after the sphericalization. We estimate the $1-\sigma$
errorbars for the anisotropy from the 10 sub-samples drawn from each
stellar halo.

\subsection{Mock halos with different shapes}
\label{sec:shapes}
We also generate smooth distribution of particles within non-spherical
regions using the Monte Carlo technique. We consider the following
three shapes for the mock stellar halos: (i) oblate spheroid $(a=1,
b=1,c=0.75)$, (ii) prolate spheroid $(a=1, b=1, c=1.5)$ and (iii)
triaxial ellipsoid $(a=1.5, b=0.88, c=0.75)$. In each type of
distribution, one million particles are distributed within a volume
comparable to the volume of the randomly extracted versions ($R=200$
kpc) of the original halo. The density distribution in each halo
follows a power law of the form $r^{-3}$. We simulate $10$ mock
stellar halos for each shape. We use these mock halos to calculate the
anisotropies introduced by the shape of the stellar halo. Here, we
would like to mention that the principal axes of the non-spherical
halos coincide with the coordinate axes. One can also consider
non-spherical halos with different orientations. We do not consider
such halos in this work for simplicity. It is worthwhile to mention
here that the coordinate system of the Bullock \& Johnston model is
oriented such that the $z$ axis is aligned along the minor axis of the
disk potential.

\subsection{Information entropy based anisotropy measures}

\subsubsection{Whole-sky anisotropy within a given radius}
\label{sec:rad}
We use the information theoretic definition of anisotropy proposed in
Pandey (2016) \citep{pandey16a} to study the anisotropy in a stellar
halo. We would like to study the anisotropy in the distribution of
stars with respect to an observer located at the centre of the
halo. We first divide the $4\,\pi$ steradian solid angle around the
central observer into a number of solid angle bins of equal size. This
is achieved by binning $\cos\theta$ and $\phi$ into $m_{\theta}$ and
$m_{\phi}$ bins respectively. This results into a total
$m_{total}=m_{\theta}m_{\phi}$ solid angle bins of equal size
$d\Omega=\sin\theta d\theta d\phi$. Each of these solid angle bins
would cover exactly the same volume $dv=\frac{r^3}{3}d\Omega$ within a
given radius $r$ from the centre of the halo. The whole sphere has a
solid angle of $4\, \pi$ steradian which is $\sim 41253$ square
degree. The solid angle or the area corresponding to each bin can be
simply obtained by dividing these numbers by $m_{total}$. Each bin
covers exactly same area on the sky. The radius $r$ can be varied
within a range $0<r\leq r_{max}$.

We define a random variable $X_{\theta\phi}$ as the event of finding a
star within a distance $r$ from the centre of the halo. There are a
total $m_{total}$ volume elements around the central observer and a
randomly chosen star can occupy any one of these volume elements. So
there are $m_{total}$ probable outcomes for this event. The
probability of the $i^{th}$ outcome is given by,
$f_{i}=\frac{n_{i}}{\sum^{m_{total}}_{i=1} \, n_{i}}$ with the
constraint $\sum^{m_{total}}_{i=1} \, f_{i}=1$. Here, $n_{i}$ is the
number count of stars in the $i^{th}$ volume element within a given
radius $r$.

The information entropy associated with the random variable
$X_{\theta\phi}$ can be expressed as,
\begin{eqnarray}
H_{\theta\phi}(r)& = &- \sum^{m_{total}}_{i=1} \, f_{i}\, \log\, f_{i} \nonumber\\ &=& 
\log N - \frac {\sum^{m_{total}}_{i=1} \, n_i \, \log n_i}{N}
\label{eq:shannon2}
\end{eqnarray}
where $N$ is the total number of stars within radius $r$. The base of
the logarithm can be chosen arbitrarily and we choose it to be $10$.

If we consider an isotropic distribution of stars in the halo then
each of the $m_{total}$ volume element would contain nearly the same
number of stars within them. The probability $f_{i}$ for each of the
$m_{total}$ volume element would be same for such a distribution. This
maximizes the uncertainty in $X_{\theta\phi}$ since each of the volume
element hosts the same number of stars. The maximum information
entropy for a given choice of $m_{\theta}$, $m_{\phi}$ and $r$ is
$(H_{\theta\phi})_{max}=\log \, m_{total}$. We define the whole-sky
anisotropy parameter within a given radius $r$ as
$a_{\theta\phi}(r)=1-\frac{H_{\theta\phi}(r)}{(H_{\theta\phi})_{max}}$.
Ideally one would expect $H_{\theta\phi}=\log \, m_{total}$ or
$a_{\theta\phi}=0$ for an isotropic distribution. However, a
fluctuations in the number counts would be present at smaller radii
due to the Poisson noise. Such fluctuation due to the Poisson noise is
expected to diminish with increasing number counts at larger
radii. However, the anisotropy parameter would never become exactly
zero for even an isotropic distribution due to the finite number of
particles and volume. A distribution can be regarded as isotropic
provided the observed anisotropies are consistent with the expected
Poisson noise at the same sampling rate and choice of binning.

The stellar halos are known to host a wealth of substructures. The
presence of these substructures is expected to introduce large
fluctuations in the star counts across the $m_{total}$ volume
elements. Any deviation from sphericity of the halo is also expected
to introduce significant fluctuations in the number count of stars.
The fluctuations in the star counts would decrease the uncertainty in
$X_{\theta\phi}$ by reducing the information entropy $H_{\theta\phi}$
associated with it. The anisotropy parameter $a_{\theta\phi}$ would be
higher for greater reduction in the information entropy. The
anisotropy $a_{\theta\phi}$ would reach its maximum value $1$ for a
distribution where all the stars occupy a single volume element
without any uncertainty ($H_{\theta\phi}=0$) in their locations.  This
extreme situation corresponds to the maximum anisotropy in the
distribution.

One can study the whole-sky anisotropy parameter $a_{\theta\phi}(r)$
by progressively probing a larger distance ($r$) from the centre of
the halo. The radial distance limit is increased in steps from a small
radius $r$ to a maximum radius $r_{max}$ for a given choice of
$m_{\theta}$ and $m_{\phi}$. Here we would like to mention that
$a_{\theta\phi}(r)$ can be also calculated in a series of independent
radial bins.

The $a_{\theta\phi}(r)$ is expected to steadily decrease with $r$ for
a homogeneous, isotropic random distribution of stars. We use
$a_{\theta\phi}(r)$ to study the radial variation of the whole-sky
anisotropy in a stellar halo.

\subsubsection{Anisotropy at fixed polar angle within a given radius}
\label{sec:pol}
One can define the anisotropy at fixed polar angle for a given choice
of $m_{\theta}$, $m_{\phi}$ and $r_{max}$ in a similar manner. The
information entropy for a fixed radius $r$ and a given $\theta$ can be
written as,
\begin{eqnarray}
H_{\phi}(\theta)& = &- \sum^{m_{\phi}}_{i=1} \, f_{i}\, \log\, f_{i} \nonumber\\ &=& 
\log N_{\theta} - \frac {\sum^{m_{\phi}}_{i=1} \, n_i \, \log n_i}{N_{\theta}}
\label{eq:shannon3}
\end{eqnarray}
Here $N_{\theta}=\sum^{m_{\phi}}_{i=1} \, n_{i}$ is the total number
of stars lying in the $m_{\phi}$ volume elements within a radial
distance $r_{max}$ at any given $\theta$. We define
$f_{i}=\frac{n_{i}}{N_{\theta}}$ in \autoref{eq:shannon3} where the
sum is carried out over all the $\phi$ bins.

We define the anisotropy parameter at fixed polar angle
$a_{\phi}(\theta)=1-\frac{H_{\phi}(\theta)}{(H_{\phi})_{max}}$ where
$(H_{\phi})_{max}=\log \, m_{\phi}$. For a fixed radial distance limit
$r_{max}=R$, we measure the anisotropy parameter $a_{\phi}(\theta)$ at
different values of $\theta$. This measures the anisotropy in the
polar direction by considering all the $\phi$ bins at each $\theta$.

\subsubsection{Anisotropy at a fixed azimuthal angle within a given radius}
\label{sec:azi}
Similarly, we define the anisotropy at fixed azimuthal
  angle for a given choice of $m_{\theta}$, $m_{\phi}$ and
$r_{max}$. The information entropy for a fixed radius $r$ and a given
$\phi$ is obtained by carrying out the sum across all the $\theta$
bins and is defined as,
\begin{eqnarray}
H_{\theta}(\phi)& = &- \sum^{m_{\theta}}_{i=1} \, f_{i}\, \log\, f_{i} \nonumber\\ &=& 
\log N_{\phi} - \frac {\sum^{m_{\theta}}_{i=1} \, n_i \, \log n_i}{N_{\phi}}
\label{eq:shannon4}
\end{eqnarray}
Here $N_{\phi}=\sum^{m_{\theta}}_{i=1} \, n_{i}$ is the total number
of stars lying in the $m_{\theta}$ volume elements within a given
$r_{max}$ at a given $\phi$. We have $f_{i}=\frac{n_{i}}{N_{\phi}}$
and the sum is carried out over all the $\theta$ bins in the
\autoref{eq:shannon4}.

We define the anisotropy parameter at fixed azimuthal angle
$a_{\theta}(\phi)=1-\frac{H_{\theta}(\phi)}{(H_{\theta})_{max}}$ where
$(H_{\theta})_{max}=\log \, m_{\theta}$. We determine the anisotropy
parameter $a_{\theta}(\phi)$ for different values of $\phi$ while
keeping the radial distance limit $r_{max}=R$ fixed. The parameter
$a_{\theta}(\phi)$ thus measures the anisotropy in the azimuthal
direction by considering all the $\theta$ bins at each $\phi$.

It may be noted that both $a_{\phi}(\theta)$ and $a_{\theta}(\phi)$
can be measured for different limiting radius and hence can be also
written as, $a_{\phi}(\theta,r)$ and $a_{\theta}(\phi,r)$
respectively.

All three anisotropy parameters defined here probe the structure of
the stellar halo and complement each other.

\begin{figure*}[htbp!]
\centering \includegraphics[width=10cm]{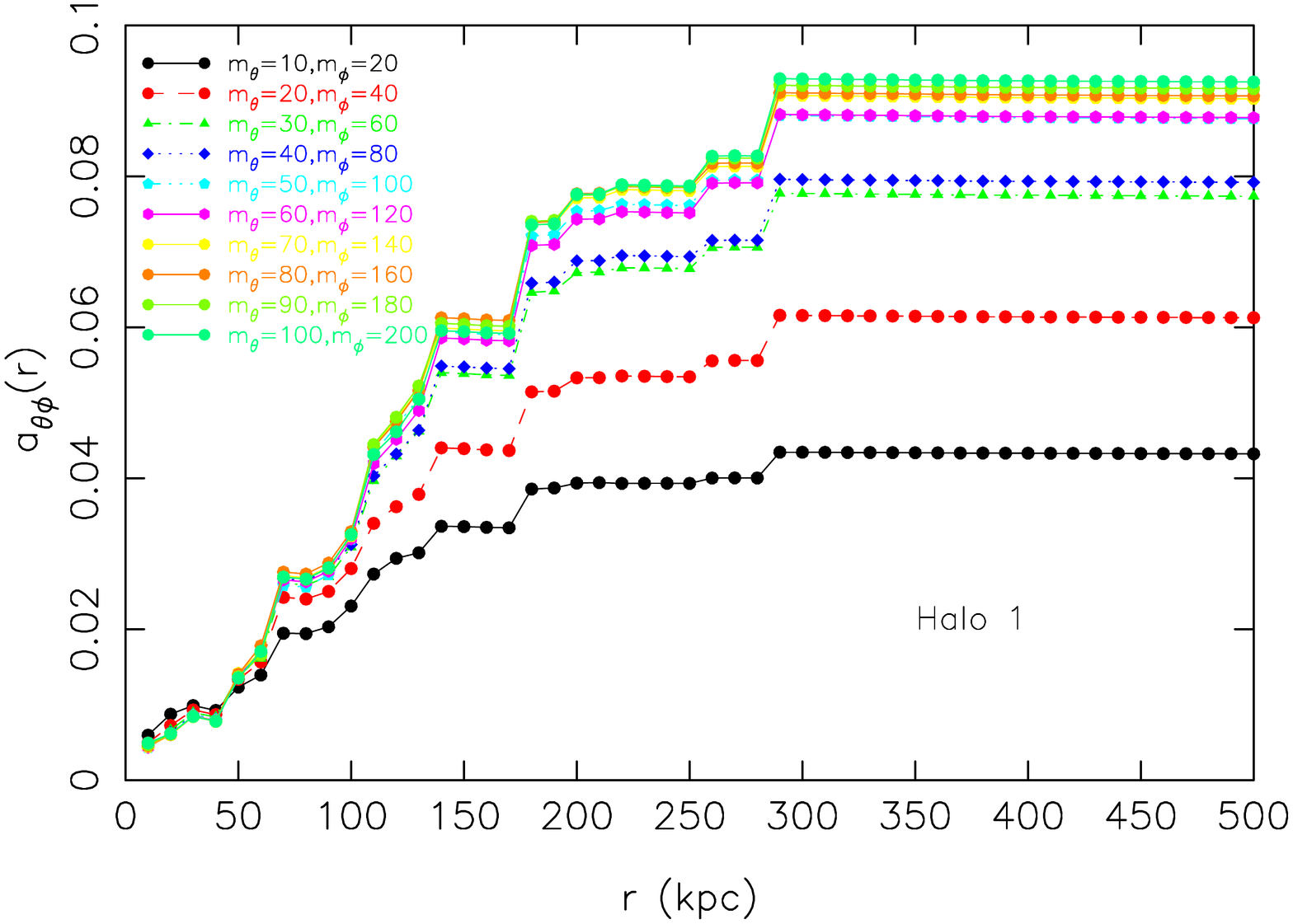}
\centering \includegraphics[width=10cm]{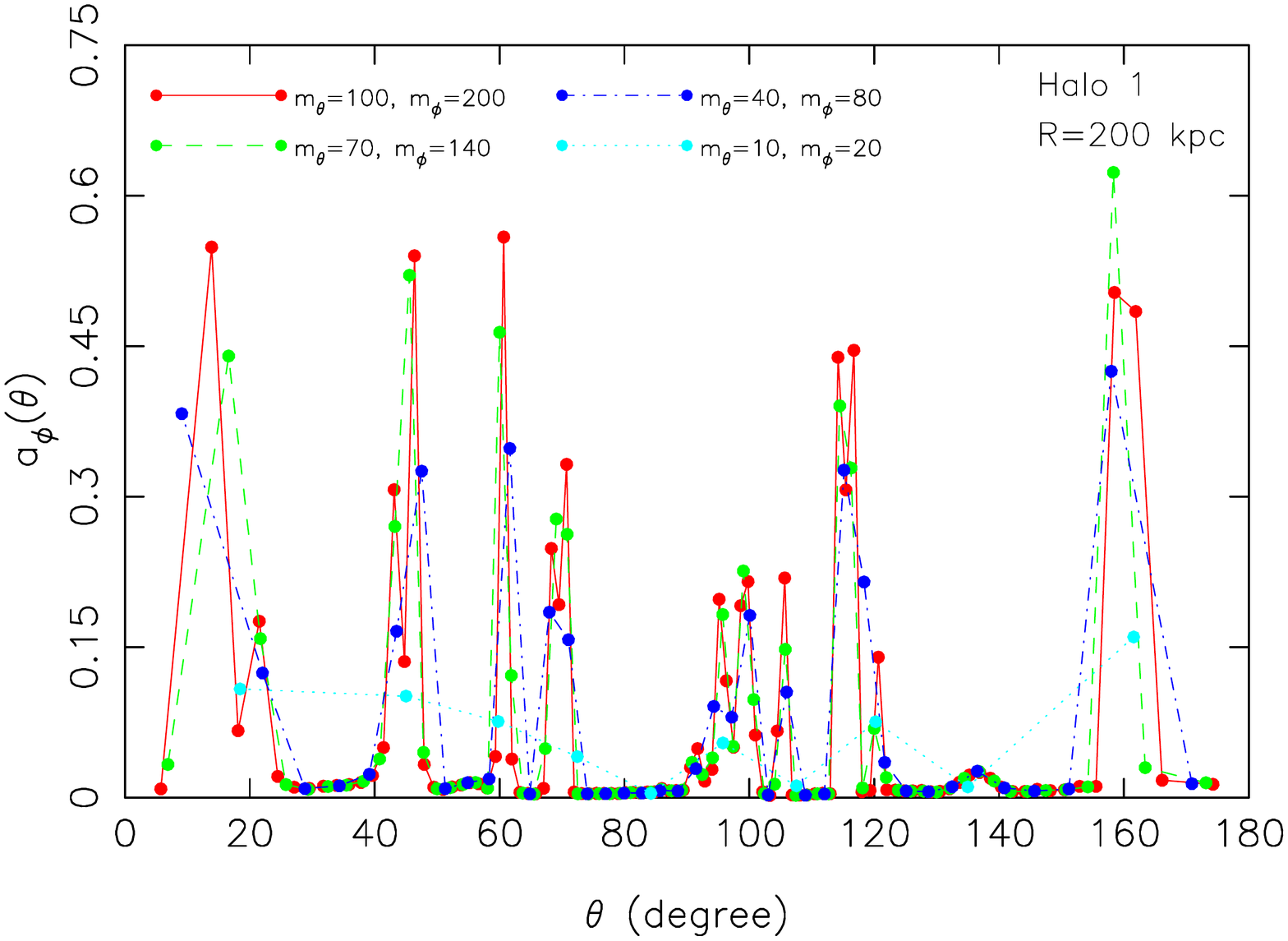}
\centering \includegraphics[width=10cm]{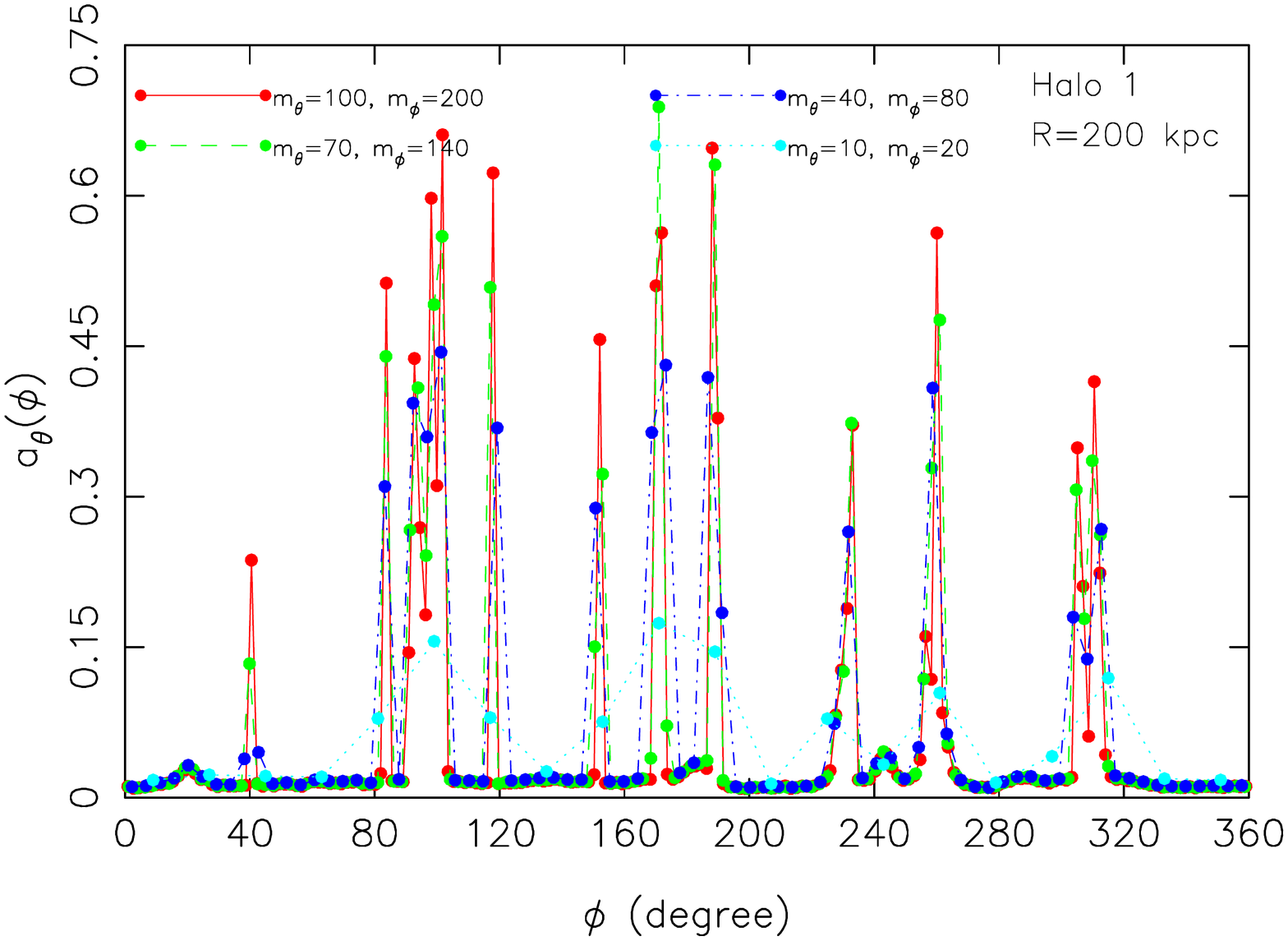}
\caption{The top, middle and bottom panels of this figure respectively
  show the whole-sky anisotropy, the anisotropy at fixed polar angle
  and anisotropy at fixed azimuthal angle in Halo 1 for different
  choices of number of bins.}
\label{fig:bins}
\end{figure*}

\section{Results}
\subsection{Effects of binning on the anisotropy}
\label{sec:binning}
The information entropy is known to be sensitive to the choice of
bins. Naturally, the anisotropy measures based on the information
entropy would be also sensitive to the binning adopted in the
analysis. However, this does not pose any problem provided we
understand the effects of binning on the anisotropy and any
comparisons of anisotropy are carried out with precisely the same
number of bins.

In this subsection, we first address the effects of bin size in the
measurement of anisotropies in the stellar halo. In
\autoref{fig:bins}, we show the different anisotropies in one of the
stellar halos (Halo 1) for different choices of $m_{\theta}$ and
$m_{\phi}$.

The top panel of \autoref{fig:bins} shows the whole-sky anisotropy in
the Halo 1 as a function of the distance $r$ from the centre of the
halo. The results are shown for $10$ different choices of bin size. We
find that the anisotropy in the stellar halo increases with the
distance from the centre of the halo. The anisotropy keeps increasing
up to a radius of $\sim 270$ kpc beyond which it plateaus to a
constant value. A larger $m_{total}$ corresponds to smaller size of
the solid angle bins. The volume covered by each solid angle bin at a
distance $r$ would decrease with increasing $m_{total}$. So the larger
values of $m_{\theta}$ and $m_{\phi}$ would result into smaller volume
elements giving rise to a larger fluctuations in the star counts. At
any given radius $r$, a larger degree of anisotropy is expected for a
larger $m_{total}$. Such differences are caused by (i) presence of
lumps and tidal streams on smaller scales and (ii) increase in the
Poisson noise due to the smaller number counts. A smaller $m_{total}$
would probe larger volume elements, thereby decreasing the Poisson
noise. It would also decrease the disparity in the abundance of lumps
and streams across the different elemental volumes. So the observed
anisotropy at any given radius $r$ should decline with the decreasing
number of solid angle bins ($m_{total}$). Interestingly, we find that
the stellar halo exhibit a minimal degree of anisotropy within a
radius of $\sim 50$ kpc. The observed anisotropies within $50$ kpc for
the different choices of $m_{\theta}$ and $m_{\phi}$ are nearly
indistinguishable. This suggests that the core of the halo has a very
smooth structure with little to no substructures. The finite
anisotropy at the core of the halo may arise due to the discreteness
noise and its anisotropic shape at these radii. The gradual increase
in the anisotropy $a_{\theta\phi}(r)$ beyond $50$ kpc indicates that
the abundance of substructures and anisotropy in their distribution
increase with the radial distance from the centre of the halo. The
variation of anisotropy ceases beyond $\sim 270$ kpc indicates the
absence of substructures beyond this radius. All the different choices
of $m_{\theta}$ and $m_{\phi}$ reveal the same trend in the variation
of $a_{\theta\phi}(r)$. We also note that whole-sky anisotropy
$a_{\theta\phi}(r)$ tends to converge at higher values of $m_{total}$.

We show the anisotropy at fixed polar angle $a_{\phi}(\theta)$ as a
function of $\theta$ and the anisotropy at fixed azimuthal angle
$a_{\theta}(\phi)$ as a function of $\phi$ in the middle and bottom
panels of \autoref{fig:bins} respectively. In both cases, we fix the
radial distance limit $r_{max}$ to $200$ kpc and show the results for
only four different choices of bins for clarity. The two anisotropy
profiles of the Halo 1 show a number of distinct spikes. These spikes
indicate a sudden rise in the anisotropy in the halo along certain
polar and azimuthal directions. The anisotropy at the spikes are
significantly larger compared to the remaining directions. These
spikes in the two anisotropy profiles indicate the presence of several
substructures in the halo.

We note that $a_{\phi}(\theta)$ and $a_{\theta}(\phi)$ are also
sensitive to the choice of the number of bins. However, the locations
of the spikes in the two anisotropy profiles do not depend on the
choice of $m_{\theta}$ and $m_{\phi}$. So the most anisotropic
directions in the stellar halo can be correctly identified with any
choice of $m_{\theta}$ and $m_{\phi}$. The degree of anisotropy at
each spike generally increases with the increasing number of bins
($m_{total}$) for the same reasons mentioned earlier.

\begin{figure*}[htbp!]
\centering \includegraphics[width=10cm]{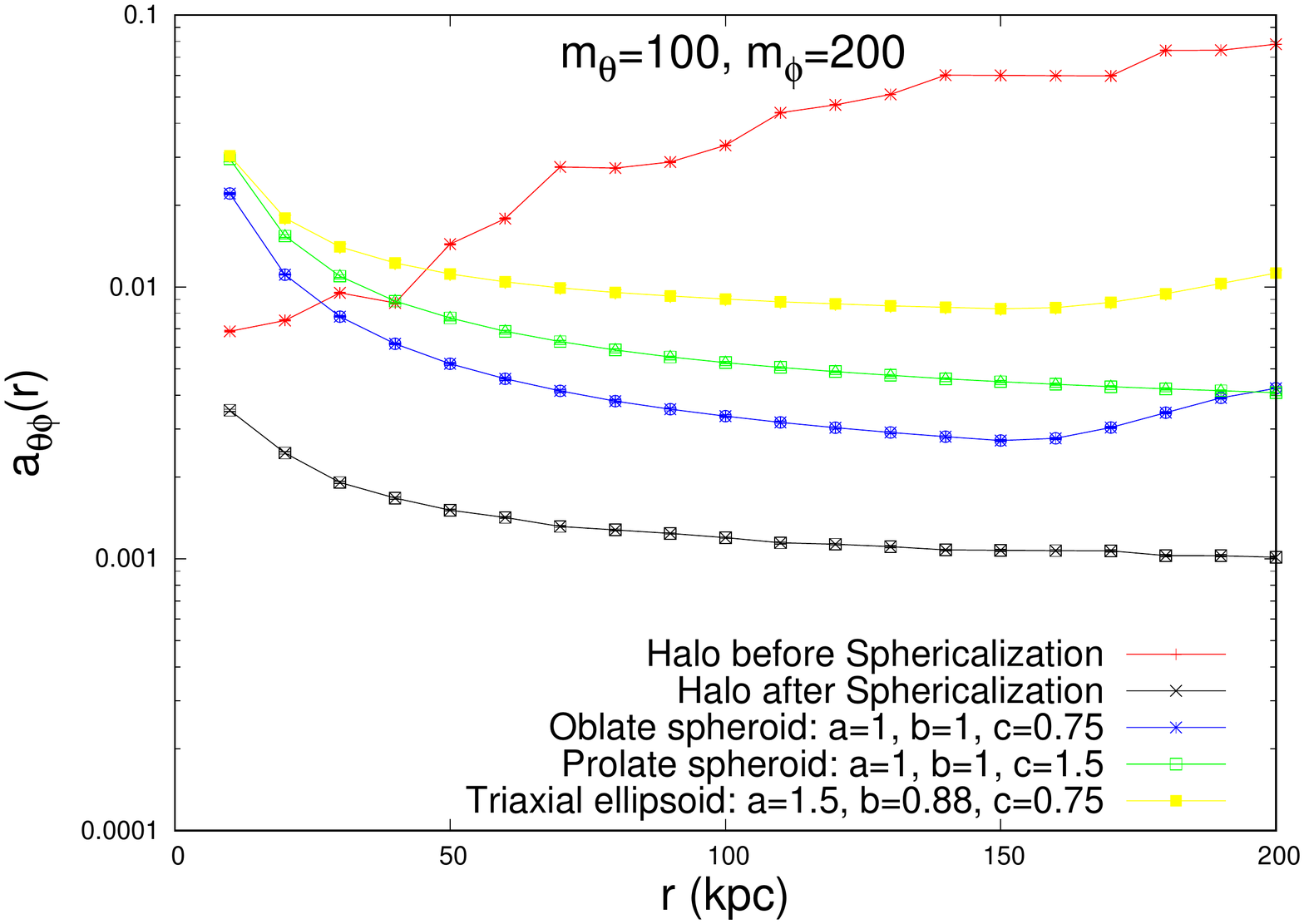}
\centering \includegraphics[width=10cm]{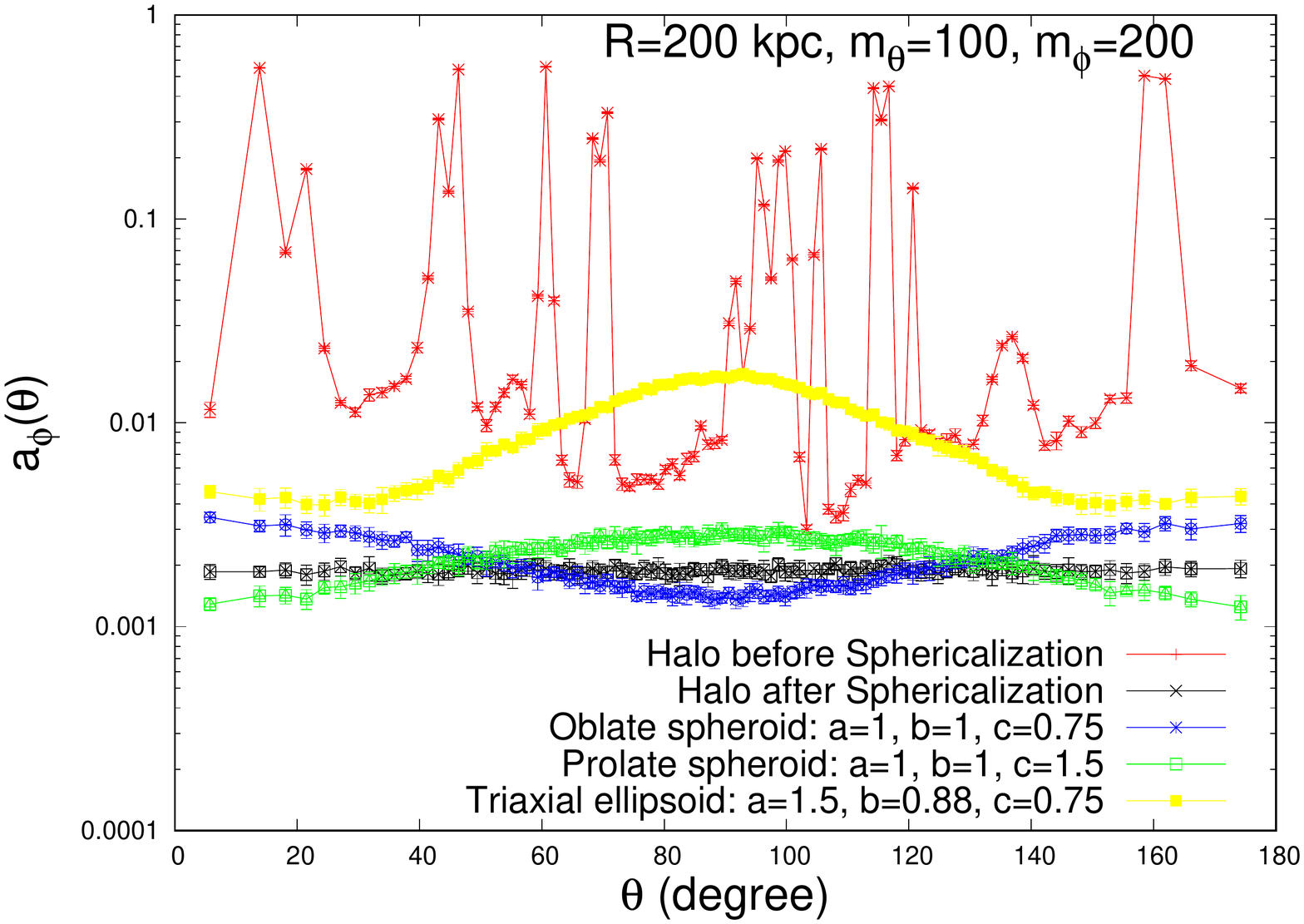}
\centering \includegraphics[width=10cm]{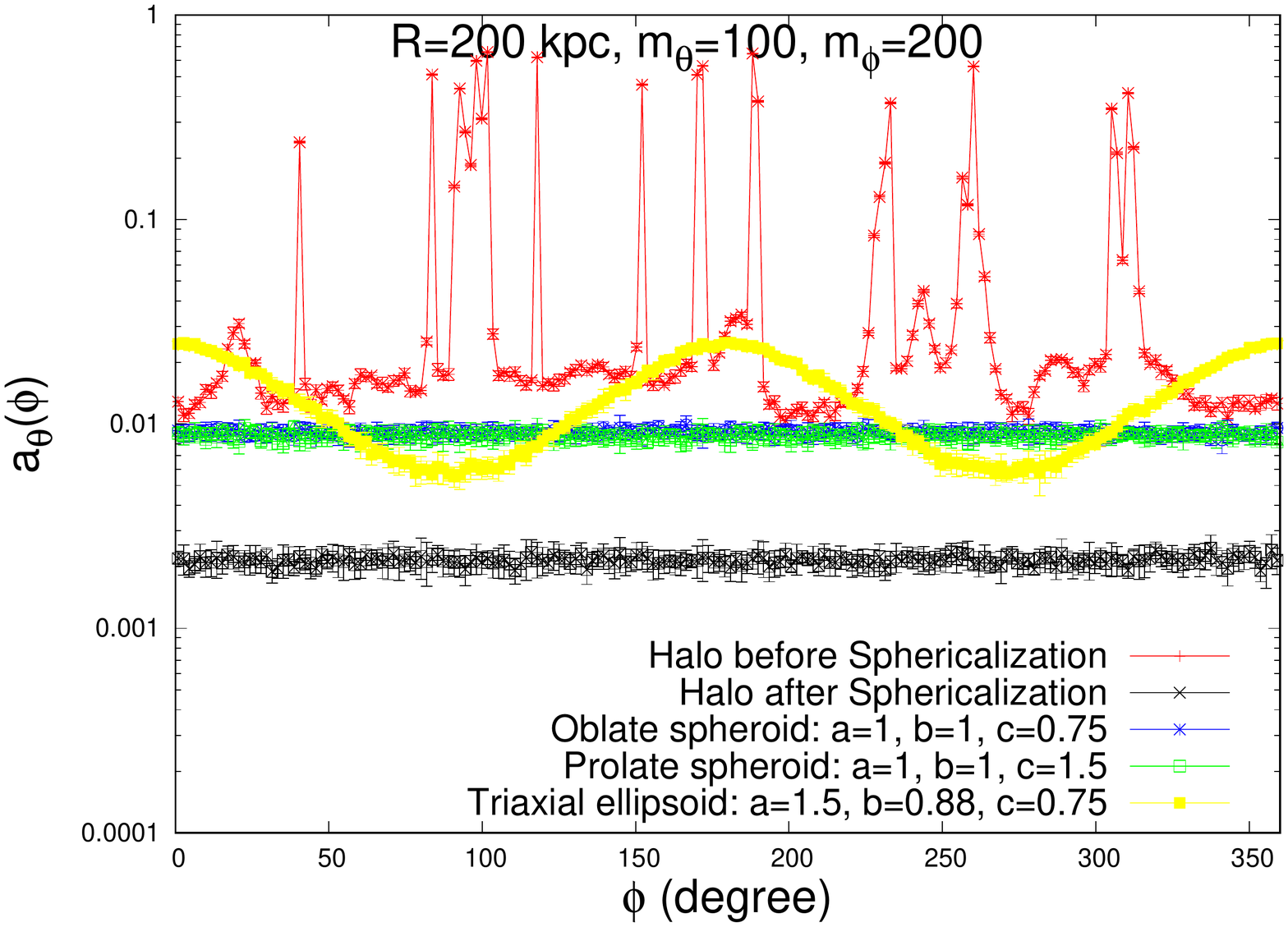}
\caption{The top, middle and bottom panels of this figure respectively
  compares the whole-sky anisotropy, anisotropy at fixed polar angle
  and anisotropy at fixed azimuthal angle in Halo 1 before and after
  the sphericalization. The anisotropies in a set of mock halos with
  same density profile but different shapes are also compared in the
  same panels. The $1-\sigma$ error-bars at each data point are
  obtained from 10 jackknife samples for the original halo and their
  sphericalized versions. We use 10 independent realizations to obtain
  the $1-\sigma$ error-bars for the mock halos.}
\label{fig:sphericalize}
\end{figure*}

\subsection{Comparison with sphericalized halos: shape versus substructures}
\label{sec:spherical}
The anisotropy in a stellar halo may arise from three different
sources: the substructures, the shape and the discreteness noise in a
stellar halo. We want to estimate their contributions to the
anisotropy in a stellar halo. We carry out an analysis following the
procedure outlined in \autoref{sec:randomize}.

We randomize the polar and azimuthal coordinates of each stellar
particle within a given radius keeping their radial distances fixed. We
refer to this process as sphericalization of the stellar halo. The
sphericalized versions of the stellar halo would not contain any
substructures. Further, all the information about the halo shape is
also lost after the sphericalization. The sphericalized halos are
expected to have the same radial density profile as the original
stellar halo.

The sphericalization of a stellar halo is thus expected to remove all
the anisotropies caused by the substructures and non-spherical shape
of the halo. Consequently, we expect a significant reduction in the
anisotropy of the stellar halo after its sphericalization. A small
residual anisotropy would still be present in the halo after the
sphericalization. This anisotropy is due to the Poisson noise that
originates from the discrete nature of the distribution. We can
compare the anisotropy in the original stellar halo and its
sphericalized versions to extract important information about the
substructures and shape of the halo.

The analysis in \autoref{sec:binning} shows that the anisotropies tend
to converge at higher values of $m_{total}$. We set $m_{\theta}=100$
and $m_{\phi}=200$ as the default choice of binning for the rest of
our analysis. This choice corresponds to an area of
$\frac{41,253}{20,000} \approx 2$ square degrees for each bin. We
compare the whole-sky anisotropy in the original halo with its
sphericalized versions in the top panel of
\autoref{fig:sphericalize}. We observe a large reduction in the
whole-sky anisotropy throughout the radius range. The non-zero
whole-sky anisotropy in the sphericalized versions of the halo is
caused by the Poisson noise. It decreases with the increasing radius
of the halo. In contrast, the whole-sky anisotropy in the original
halo rises with increasing radius. The ratio of anisotropies in the
original and the sphericalized halos increases from 2 to 80 in the
radius range $10$ kpc to $200$ kpc. The increase in amplitude of the
ratio with radius indicates that the stellar halo becomes
progressively more anisotropic at larger radii. We compare the
whole-sky anisotropies in a set of smooth halos with different shapes
in the same panel and find that the whole-sky anisotropies are
significantly larger in these halos than the sphericalized versions of
the original one. They even surpass the anisotropies in the original
halo at a radius $<50$ kpc. However, the whole-sky anisotropies in
oblate, prolate and triaxial halos are significantly smaller than the
original halo at a radius beyond $50$ kpc. It indicates that the
whole-sky anisotropies at smaller and larger radii may be respectively
determined by the shape and substructures of the halo.

In \autoref{fig:sphericalize}, we note that the anisotropies at fixed
polar angle or fixed azimuthal angle in the original stellar halo have
a smooth and an irregular component. The smooth and irregular
components are not separately visible in the whole-sky anisotropies as
it is calculated at each radius using all the solid angle bins. The
irregular component originates from the substructures on smaller
scales and is expected to disappear after the sphericalization. The
smooth component of the anisotropy must remain unchanged after the
sphericalization provided it originates from the discreteness noise
alone. However, there could be an additional contribution to
anisotropy from the non-spherical shape of the halo. We expect a
reduction in the anisotropy of the smooth component if the shape of
the halo contributes to the anisotropy at fixed polar angle and
anisotropy at fixed azimuthal angle. We show these anisotropies inside
Halo 1 in the middle and bottom panels of
\autoref{fig:sphericalize}. In both the panels, we find a significant
reduction in the anisotropies after the sphericalization. The
sphericalization of the halo completely eliminates the highly
irregular component of the anisotropy. Furthermore, it also
significantly depletes the approximately uniform component of the
anisotropy in the original halo. The residual anisotropy in the halo
after the sphericalization can only arise due to the discreteness
noise.  We find that the residual anisotropy in the sphericalized
halos are $\sim 4-5$ times smaller than the approximately uniform
component of the anisotropy in the original halo. So the approximately
uniform component of anisotropy in the original halo can not arise due
to the discreteness noise alone. On the other hand, the fluctuating
part of the anisotropy in the halo is $\sim 200-300$ times larger than
the residual anisotropy. It indicates that the substructures are the
most dominant source of anisotropy in the stellar halo.

These results suggest that the substructures and the discreteness
noise may not be the only sources of anisotropy in a stellar halo. The
shape of the stellar halo may also significantly contribute to the
anisotropy. We can quantify the contribution to the anisotropy from
the halo shape. We compare the results from the original halo with a
set of smooth halos with the same density profile ($\sim r^{-3}$) but
different shapes (\autoref{sec:shapes}). We find that the
non-spherical shape of a stellar halo can indeed introduce additional
anisotropies in addition to the anisotropies due to the discreteness
noise. We note that none of the three shapes considered here
accurately describe the shape of the Halo 1. Nonetheless, our results
indicate that the reduction in the approximately uniform component of
anisotropy after sphericalization could be explained by the
non-spherical shape of the halo. This distinction between the
anisotropies may allow us to constrain the shape of the stellar halo
as well as its substructure content.

The different orientations of the non-spherical halos would inevitably
modulate the signals from the large-scale asymmetry present in the
halo. We do not address the possible degeneracy between shape and
orientation of the low-order ellipsoidal contribution in this work. We
plan to address these issues in a future work.

\begin{figure*}[htbp!]
\centering \includegraphics[width=14cm]{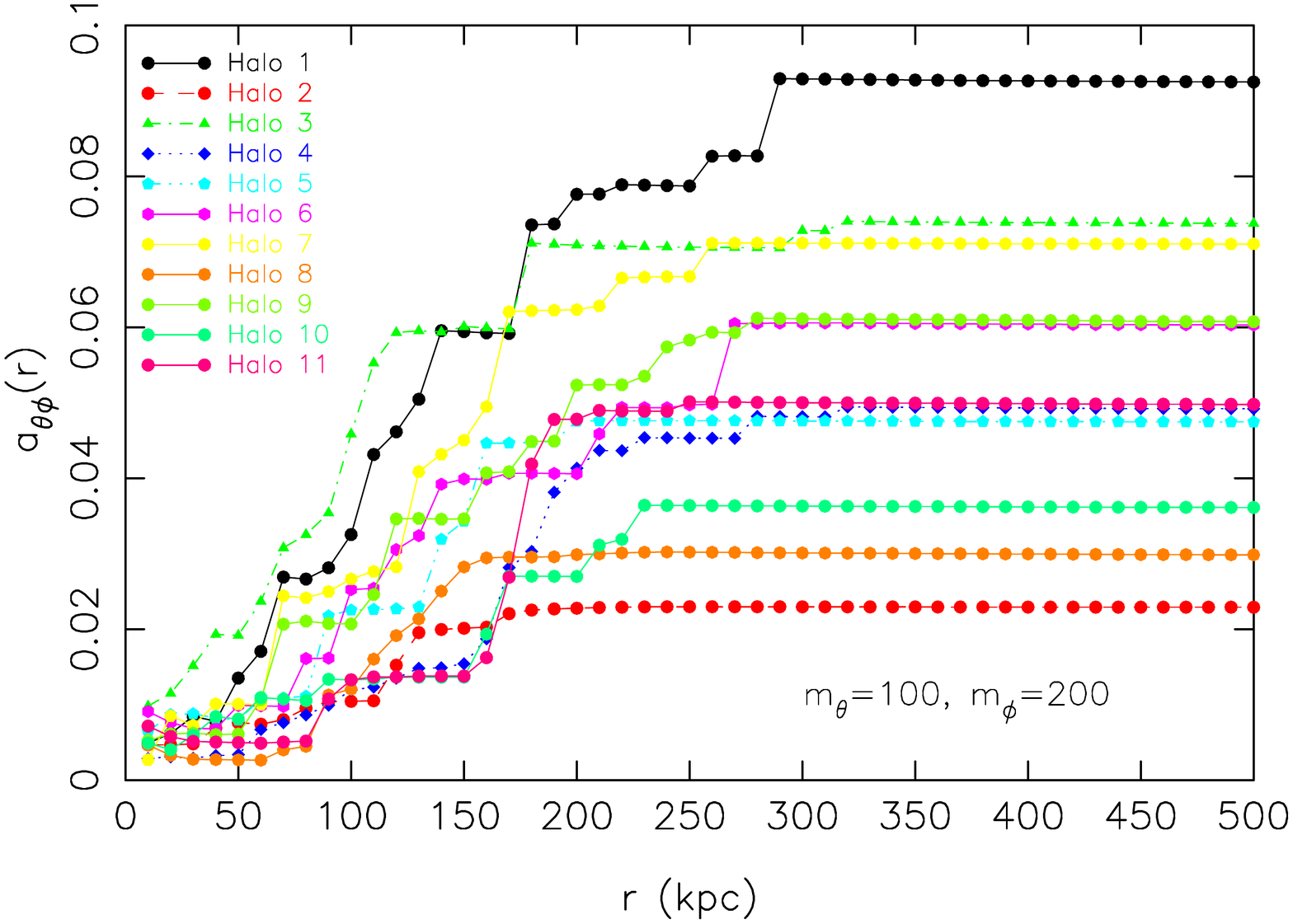}
\caption{This figure compares the whole-sky anisotropy as a function
  of the distance from the centre of the halo for all the 11 halos in
  the Bullock \& Johnston suite of simulations.}
\label{fig:all}
\end{figure*}

\subsection{Comparison of anisotropy in different stellar halos}

We analyze all the 11 stellar halos from the Bullock \& Johnston suite
of simulations. The whole-sky anisotropy in these 11 stellar halos are
compared with each other in \autoref{fig:all}. The results suggest
that the different stellar halos exhibit different degrees of
whole-sky anisotropy at any given radius. The radial variation of the
anisotropy is also different in each halo. The anisotropy plateaus out
at different radii for each stellar halo. For instance, the whole-sky
anisotropy increases up to 150 kpc in Halo 2, whereas it keeps
increasing up to 270 kpc in Halo 1. Such differences in the anisotropy
may arise from the diverse assembly history of the stellar
halos. Despite such differences, nearly all the stellar halos have a
tiny and similar anisotropy within a radius of 50 kpc from the
centre. This re-emphasizes that all the stellar halos have a smoother
core. The gradual increase of the anisotropy in the radially outward
direction in all cases indicates an `inside out' build-up of the
stellar halo. The simulations used in this work assume
  hierarchical structure formation and the results of our analysis are
  qualitatively consistent with this assumption. A more detailed
  analysis is required to test the standard view of how mass is
  deposited in the stellar halo \citep{amorisco}.

%Here, we do not show any other plots to compare the anisotropies
%\textcolor{red}{at fixed polar or azimuthal angle} in different
%halos. These anisotropies exhibit the same trends as discussed for
%Halo 1 in \autoref{sec:spherical} and \autoref{sec:mapping}. We count
%the maximum number of peaks observed in the two anisotropy profiles of
%the halos and compare them with the number of surviving satellites in
%the respective halos as listed in Table 1 of \citep{bullock05}. We
%find that these numbers are very close in most cases, implying that
%each peak correspond to an individual substructure. However, this may
%not be always true. We need to exercise caution particularly for the
%peaks that change amplitude with the increasing radial distance.

\section{Conclusions}

In this work, we analyze the anisotropy in a set of stellar halos from
the Bullock \& Johnston suite of simulations. Our main findings can be
summarized as follows:\\

(i) All the stellar halos exhibit a very smooth structure within a
radius of $\sim 50$ kpc from their centres. The anisotropy in each
halo increases with the radial distance from the centre and plateaus
out beyond a certain radius. The halos are most anisotropic in their
outer parts. The degree of anisotropy and its radial variation is
different for each stellar halo suggesting a wide variety of assembly
history. Such differences may provide useful constraints on the
late-time accretion histories of galaxies. Our results indicate an
inside out formation of stellar halo which is qualitatively consistent
with the assumption of hierarchical structure formation in the
simulations.\\

(ii) The anisotropies at fixed polar or azimuthal angle in a stellar
halo show two distinct contributions. These anisotropies can be
thought of as a superposition of a nearly constant part and a highly
fluctuating part. The nearly constant part represents the anisotropy
in the smooth component of the halo, whereas the fluctuating part
corresponds to the anisotropy from its lumpy component. We
sphericalize the stellar halos by randomizing the polar and azimuthal
coordinates of all the stellar particles and keeping their radial
coordinates fixed. This destroys all the substructures and hence
eliminates the fluctuating part of the anisotropy. The approximately
uniform component of the anisotropy should not be affected by the
sphericalization if the anisotropy in the smooth component originates
purely from the Poisson noise. However, we observe a reduction in the
approximately uniform component of the anisotropy after the
sphericalization, which suggests a non-spherical shape of the stellar
halo. A non-spherical shape of the halo may introduce additional
anisotropies in the distribution. Any information about the
non-spherical shape of the original halo would be completely wiped out
after the sphericalization. We note that the reduction in the
approximately uniform component of the anisotropy may be explained by
considering different shapes of the stellar halo. It implies that the
reduction in the approximately uniform component of the anisotropy is
caused by the destruction of the shape of the halo, and the
destruction of the substructures eliminates the fluctuating part. In
other words, the anisotropy signal includes contributions from both
the large-scale non-spherical shape of the halo and the small scale
substructure, with the latter dominating separate the contributions to
anisotropy from the shape and substructure.\\

%(iii) We show that one can map the radial distributions of the
%individual substructures by \textcolor{red}{combining the different
%  anisotropy profiles of the halo}. The maximum number of peaks
%observed in the anisotropy profiles \textcolor{red}{at fixed polar or
%  azimuthal angle} in each halo is nearly the same as the number of
%surviving satellites in that halo. It indicates that most of the peaks
%may be associated with individual substructures.\\

We propose a statistical method to quantify the substructures and
shape of the stellar halo. The limitations of the method are the
following. The entropy depends on the choice of binning and the
Poisson sampling rate. However, the relative character of the entropy
does not pose any problem provided the distributions are compared with
the same binning and sampling rate. Further, the existing
observational datasets for galactic stellar halo do not provide a full
sky coverage. One also needs to correct for the selection biases and
the systematic present in the observational data before applying the
proposed method to constrain the shape of the stellar halo and the
substructures therein.

The shape of the stellar halo is usually determined directly by
fitting prolate, oblate and triaxial smooth and broken power-law
models \citep{bell08, deason11}. One can also infer the shape by
analyzing the orbits of the stars and the dynamical modelling of the
stellar halo \citep{bovy16, sato22}. Our method is different from
these traditional techniques. It is solely based on the measurement of
anisotropies present in the halo. The method involves only a
sphericalization of the stellar halo and a comparison of the
anisotropies in the halo before and after the sphericalization. 

The method is entirely based on the measurement of the spatial
anisotropy. It does not require any dynamical modelling and the
analysis of the stellar orbits. So the velocities of the stellar
particles are not needed. It also does not require us to identify the
individual substructures in the stellar halo. The method is relatively
simple and can serve as an alternative and complementary technique for
measuring shape of the stellar halo.

It would also be interesting to relate the anisotropy measurements
with the stellar metallicity distribution and chemical abundance
patterns of the halo. The distribution and abundance of the metals in
the halo are crucial for understanding the chemical evolution of the
galaxy \citep{freeman02}. The chemical evolution of the halo can be
predicted in some models \citep{robertson05, font06} and can be also
studied from observations. The old stellar population in the halo that
formed earlier are known to be metal-poor \citep{eggen62,
  robertson05}. These metal-poor stars are the best tracers for the
assembly history of the galaxy. They are usually found at larger
distances from the centre of the halo. Contrary to this, the
metal-rich stars are populated near the centre of the
halo. Simulations suggest that the outer part of the stellar halo is
primarily assembled from the less-massive satellites, which are most
metal-poor due to the truncated star formation \citep{amorisco}. The
existence of such a metallicity gradient in the stellar halo is
studied in several works using simulations \citep{starkenburg,
  salvadori, tissera} and observations \citep{fernandez, beers17,
  dietz20}. We want to explore the possible relations between the
metallicity distributions and the anisotropy using simulated stellar
halos in future works.

Our method can be also applied to the distribution of stars with a
partial sky coverage. In future, we also plan to use our method to
study the stellar halo of the Milky Way using observational data from
SDSS \citep{york00} and Gaia \citep{gaia}.

\section*{ACKNOWLEDGEMENT}
I sincerely thank an anonymous reviewer for the detailed comments and
valuable suggestions that helped me to significantly improve the
presentation in the draft. The author acknowledges financial support
from the SERB, DST, Government of India through the project
CRG/2019/001110. BP would also like to acknowledge IUCAA, Pune, for
providing support through the associateship programme.

\end{document}